\documentclass[%
reprint,
twocolumns,
aps,
pre,
showpacs,
notitlepage,
nobibnotes,
superscriptaddress
]{revtex4-2}

\usepackage{float}
\usepackage{bigints}
\usepackage{psfrag}
\usepackage{grffile}
\usepackage{verbatim}
\usepackage{microtype}
\usepackage{multirow}
\usepackage{enumitem}
\usepackage{amsmath}
\usepackage{amssymb}
\usepackage{amsthm}
\usepackage{mathrsfs}
\usepackage{mathtools}
\usepackage{graphicx}
\usepackage{tikz}
\usepackage{bbm}
\usepackage{subfig}
\usepackage[linesnumbered,ruled]{algorithm2e}
\usepackage{color}
\usepackage{tcolorbox}

\usepackage{ragged2e}

\definecolor{myblue}{rgb}{0.153,0.322,0.706}
\usepackage[colorlinks,linkcolor=myblue,urlcolor=myblue,citecolor=myblue]{hyperref}
\usepackage{geometry}
\usepackage{url}
\usepackage{hyperref}
\usepackage{xcolor}

\newcommand{\be}{\begin{equation}}
\newcommand{\ee}{\end{equation}}

\def\bc{\begin{center}}
\def\ec{\end{center}}
\def\bea{\begin{eqnarray}}
\def\eea{\end{eqnarray}}

\DeclareCaptionJustification{justified}{\justifying}
\definecolor{airforceblue}{rgb}{0.36, 0.54, 0.66}
\definecolor{brickred}{rgb}{0.8, 0.25, 0.33}
\definecolor{amber}{rgb}{1.0, 0.75, 0.0}
\definecolor{applegreen}{rgb}{0.55, 0.71, 0.0}
\definecolor{magenta}{rgb}{0.965, 0, 0.859}

\newcommand{\deletetext}[1]{}

\newcommand{\ssp}{\nobreak\hspace{0.1em}}
\newcommand{\er}[1]{Eq.\ssp\eqref{#1}}

\newcommand{\era}[2]{Eqs.\ssp(\ref{#1}) and (\ref{#2})}

\newcommand{\Er}[1]{Equation\ssp\eqref{#1}}

\newcommand{\Era}[2]{Equations\ssp(\ref{#1}) and (\ref{#2})}

\renewcommand{\S}{{\cal S}}
\newcommand{\E}{{\cal E}}

\begin{document}
\title{Spectral theory for dynamical large deviations in non-Markov self-interacting processes}

\author{Francesco Coghi}
\email{francesco.coghi@leicester.ac.uk}
\affiliation{School of Computing and Mathematical Sciences, University of Leicester, Leicester, LE1 7RH, UK}
\affiliation{School of Physics and Astronomy, University of Nottingham, Nottingham, NG7 2RD, UK}
\affiliation{Centre for the Mathematics and Theoretical Physics of Quantum Non-Equilibrium Systems,
University of Nottingham, Nottingham, NG7 2RD, UK}

\author{Juan P.\ Garrahan}
\affiliation{School of Physics and Astronomy, University of Nottingham, Nottingham, NG7 2RD, UK}
\affiliation{Centre for the Mathematics and Theoretical Physics of Quantum Non-Equilibrium Systems, University of Nottingham, Nottingham, NG7 2RD, UK}

\date{\today}

\begin{abstract}
    We develop a spectral theory for dynamical large deviations in non-Markov jump processes and non-Markov chains, whose dynamics depends on the past through state- and jump-dependent empirical observables. We demonstrate that a multiscale Wentzel-Kramers-Brillouin-Jeffreys (WKBJ) Ansatz separates fast configurational relaxation from slow memory evolution, reducing the Feynman--Kac equation for occupation and flux statistics to an eigenvalue problem for a new tilted operator coupled to Hamilton--Jacobi characteristics. This provides a computationally efficient framework for quantifying fluctuations in a broad class of non-Markovian systems. We illustrate our general results with a bistable self-induced East model.
\end{abstract}

\maketitle

\paragraph*{\bf \em Introduction.} 

Small autochemotactic organisms, such as \textit{E.\ Coli}, \textit{P.\ Aeruginosa} and \textit{Pharaoh's ants}, as well as state-of-the-art artificial agents, store information in the surrounding environment~\cite{budrene1995dynamics,sumpter2003from,jackson2006longevity,howse2007self-motile,thutupalli2011swarming,reid2012slime,zhao2013psl-trails,gelimson2016multicellular}. As such, the evolving environment aids in coordinating behaviour within groups~\cite{brenner1998physical,hokmabad2022chemotactic,nakayama2023tunable}. This whole dynamics can be modelled using self-interacting processes~\cite{benaim2002self-interacting,moral2007self-interacting,schreiber2001urn-models,kurtzmann2010the-ode-method,coghi2025self-interacting}, which combine field-theoretic~\cite{keller1970initiation,tsori2004self-trapping,grima2005strong-coupling,sengupta2009dynamics,pohl2014dynamic,gelimson2015collective,kranz2016effective,grafke2017spatiotemporal} and reinforced-processes~\cite{coppersmith1986random,toth2001self-interacting,othmer2006aggregation,pemantle2007a-survey,erschler2011stuck,kious2016stuck,barbier-chebbah2022self-interacting,bremont2024exact} features. 

Self-interacting processes consist of a wide class of non-Markov systems where the memory is encoded in a process-dependent time-additive empirical observable (which mimics the environment in autochemotactic systems). Interestingly, long-range temporal correlations inherent in the time-additive memory yield rich dynamical behaviour, such as accelerated first-passage dynamics~\cite{aleksian2024self-interacting,coghi2025accelerated} and dynamical ergodicity breaking~\cite{benaim2002self-interacting,benaim2005self-interacting,coghi2024current,coghi2026proof}.

Recently, in order to study atypical behaviour, a general level-2.5 large deviation (LD) theory for self-interacting jump processes was proposed~\cite{budhiraja2025jump,coghi2026level,coghi2026proof} using a combination of stochastic control and variational techniques~\cite{budhiraja2019analysis,barato2015a-formal,chetrite2015variational}. This variational formulation is general and well-suited for extending kinetic and thermodynamic uncertainty relations to non-Markovian dynamics. However, we know from the Markovian case~\cite{touchette2009the-large,garrahan2018aspects,jack2020ergodicity} that the most efficient approach to quantifying large deviations is to study the spectral properties of the tilted operators that appear in the dual description in terms of generating functions. 

Here, we provide such dual description for  general time-additive observables of self-interacting jump processes (SIJPs) and self-interacting Markov chains (SIMCs). By deriving the corresponding Feynman--Kac equations~\cite{kac1949on-distributions,fitzsimmons1999kacs,chetrite2015nonequilibrium}, and approximating their solutions using a multiscale  Wentzel-Kramers-Brillouin-Jeffreys (WKBJ)  Ansatz~\cite{assaf2017wkb-theory} (which makes transparent a fundamental separation of time scales), we derive eigenvalue equations for a new class of tilted generators (for SIJPs) or propagators (SIMCs), thereby extending the spectral theory of large deviations from Markov~\cite{touchette2009the-large,jack2010large,chetrite2015nonequilibrium} to non-Markov self-interacting processes. As such, we recast the LD problem as the diagonalisation of a continuous family of tilted operator coupled to Hamilton-Jacobi characteristics (the latter coupling absent in the Markovian case), substantially reducing the computational cost of studying fluctuations in broad classes of non-Markovian processes, and enabling the use of efficient spectral methods for large state spaces such as tensor-network techniques~\cite{banuls2023tensor} and Krylov-subspace eigensolvers~\cite{saad2011numerical}.

\paragraph*{\bf \em Self-interacting jump processes.}
Similarly to~\cite{coghi2026level,coghi2026proof}, we consider a SIJP as the continuous-time chain $\left( X_t \right)_{0 \leq t \leq T}$ of discrete configurations, $x \in \S$ with $|\S| = d$. We denote $\E$ the set of edges or allowed transitions $(x,y)$ with $x \neq y$ (for notation and conventions see 
\hyperref[appendix:Notation]{End Matter}). 
At time $t$, the SIJP jumps from the current state $x$ to $y$, such that $(x,y) \in \E$, according to a rate $Q_{xy}(A_t) \in \mathbb{R}_+$ that depends on a general \textit{empirical observable}, a.k.a.\ \textit{memory}, of the process~\cite{touchette2009the-large,chetrite2015nonequilibrium,coghi2026proof}
\begin{equation}
    \label{eq:EmpObs}
    \begin{split}
    A_t &= t^{-1} \int_0^t f_{X_{t'}} \, d{t'} +
    t^{-1} \sum_{\substack{0 \leq t' \leq t \\ (X_{t'_-}, X_{t'}) \in \E}} 
        g_{X_{t'_-}X_{t'}} \, ,  
    \end{split}
\end{equation}
where $t'$ are jump times, and $f: \mathcal{S} \rightarrow \mathbb{R}$ and $g: \E \rightarrow \mathbb{R}$ are bounded functions. We set $A_0 = 0$ by convention. \Er{eq:EmpObs} and all subsequent evolution equations involving $t^{-1}$ are understood for $t>0$, with quantities at $t=0$ defined by their right limits (exception made for $A_0$).

For the matrix $Q(A_t)$ to be a stochastic generator, we define its diagonal components to be the negative escape rates 
\begin{equation}
    \label{eq:DiagRateSIJP}
    Q_{xx}(A_t) = - \sum_{\substack{y:  (x,y) \in \E}} Q_{xy}(A_t) \, .
\end{equation}
The dependence of the generator at time $t$ on the trajectory up to that point through \er{eq:EmpObs} makes the process non-Markovian. 


Within a time window $[t,T)$, two central objects of interest are the {\em empirical measure}
\begin{equation}
    \label{eq:EmpOccSIJP}
    L_x(t,T) \coloneqq
    (T-t)^{-1} \int_t^T \mathbf{1}_x(X_{t'}) \, 
    d{t'} 
    \, ,
\end{equation}
namely, the fraction of time that the current realisation of the process has spent in each configuration in $\S$, and the \textit{empirical flux}
\begin{equation}
    \label{eq:EmpFluxSIJP}
     \Phi_{xy}(t,T) \coloneqq (T-t)^{-1} \sum_{\substack{t \leq t' \leq T \\ (X_{t'_-}, X_{t'}) \in \E}} \mathbf{1}_{x}(X_{t_{-}'}) 
         \mathbf{1}_{y}(X_{t'}) \, ,
\end{equation}
namely, the number of jumps per unit time for each $(x,y) \in \E$. In the following, when assuming the initial time to be $t=0$, we will write $L(T)$ for empirical measure and $\Phi(T)$ for flux to keep notation compact.

We assume that for every admissible value of $A_t=a$, the frozen generator $Q(a)$ is irreducible on $\mathcal{E}$. We also assume that the map $a\mapsto Q(a)$ is at least $C^1$ and that the relevant rates are uniformly positive,  $Q_{xy}(a)\ge q_->0$ for all $(x,y) \in \mathcal{E}$, for all admissible $a$. Physically, this means that although $A_t$ changes over time, it is not allowed to break the state space in disconnected components and that a unique accompanying distribution $\gamma(a)$ exists, $\gamma(a) Q(a) = 0$, for every value of $a$. 

We also assume that there exists a compact interval $K\subset\mathbb R$ such that the self-consistency map
\begin{equation}
    F(a)
    =
    f \gamma(a)
    +
    g \gamma(a) Q(a)
\end{equation}
satisfies $F(K)\subseteq K$. Under this assumption, Brouwer's fixed-point theorem gives at least one self-consistent stationary pair for the SIJP, necessary for the large-deviation analysis below. Finally, we must impose that at least one stationary state is attracting. This implies that the empirical measure and flux converge in the long-time limit to typical values, which are not necessarily unique. Further details on these assumptions are discussed in Ref.\ssp\cite{coghi2026proof}.

In the long-time limit for a SIJP we then have
\begin{equation}
    \label{eq:TypBehav}
    L(T) \rightarrow \pi 
    \, , \;\;
    \Phi(T) \rightarrow \varphi
    \, ,
\end{equation}
as a convergence in probability for some pair $(\pi,\varphi)$ such that
\begin{equation}
    \label{eq:StatSIJP}
    \pi Q(\overline{a}) = 0 \, ,
\end{equation}
where 
\begin{equation}
    A_t \rightarrow
    \overline{a} \coloneqq 
    f \pi + g \varphi \, ,
\end{equation}
and flow conservation is satisfied for all $x \in \S$,
\begin{equation}
    \label{eq:FlowConservation}
    \sum_{y:(x,y) \in \E} \varphi_{xy }= \sum_{y:(y,x) \in \E} \varphi_{yx} \, ,
\end{equation}
The corresponding notation and definitions for the discrete time case are given below in \hyperref[appendix:SIMCs]{\em End-matter}.

\paragraph*{\bf \em Generalised Feynman--Kac dynamics
.} 
We focus on the moment generating function (MGF) of the joint observable $(L(t,T),\Phi(t,T))$ defined as
\begin{equation}
    \label{eq:Moment}
    Z_{x}(\zeta,\psi;t,a) \coloneqq \mathbb{E} \left[ e^{(T-t) B_{t,T}} | X_t = x, A_t = a \right] \, ,
\end{equation}
where
\begin{equation}
    \label{eq:ObsDualTraj}
    B_{t,T} \coloneqq 
    \zeta \, L(t,T) + \psi \, \Phi(t,T) \, .
\end{equation}
Here, $\zeta$ and $\psi$ are the counting fields for the observables $L$ and $\Phi$. 
By conditioning on the joint $(X_t,A_t)$ we are effectively describing the generating function as that of a Markov process in an extended space. In the following, we suppress the dependence on $(\zeta,\psi)$ when it is obvious from the context.

The time-extensivity of the observables means that we can write the exponent in \er{eq:Moment} as 
\begin{equation}
    (T-t)B_{t,T} = \delta t \, B_{t,t+\delta t} + (T-t-\delta t) B_{t+\delta t,T} \, ,
\end{equation}
for some time increment $\delta t$. Using conditioning and Markovianity, we can in turn express \er{eq:Moment} as
\begin{equation}
    \begin{split}
    Z_{x}(t,a) =  \mathbb{E} \left[ e^{\delta t B_{t,t+\delta t}} Z_{x'}(t+\delta t,a') | X_t = x, A_t = a \right] \, .
    \end{split}
\end{equation}
For small $\delta t$, the expectation is on trajectories with no jumps, or at most one jump that (to leading order) occurs at the end of the time interval $\delta t$. We can then write
\begin{align}
    Z_{x}(t,a) &= e^{\delta t \zeta_x} \delta t  \sum_{y: (x,y) \in \E} Q_{xy} \left( \frac{t}{t+\delta t} a + \frac{\delta t}{t+\delta t} f_x \right) \times 
    \nonumber \\ 
    &\hspace{-1.cm}  \times Z_{y} \left( t+\delta t,\frac{t}{t+\delta t} a + \frac{\delta t}{t+\delta t}\left( f_x + \frac{g_{xy}}{\delta t}\right) \right) \, +
        \nonumber \\ 
    &\hspace{-1cm}
    + e^{\delta t \zeta_x} \left( 1 - \sum_{y: (x,y) \in \E} \int_{t}^{t+\delta t} Q_{xy}(A_{t'}) \, dt' \right) \times
        \nonumber \\  
    &\hspace{-1.cm} \times Z_{x} \left( t+\delta t,\frac{t}{t+\delta t} a + \frac{\delta t}{t+\delta t}f_x \right) 
    \, .
    \label{eq:JumpOrNot}
\end{align}
In the $\delta t \to 0$ limit, \er{eq:JumpOrNot} becomes
\begin{equation}
    \label{eq:FKOh}
    \begin{split}
    0 &= \partial_t Z_{x}(t,a) + \zeta_x Z_{x}(t,a) +  \sum_{y: (x,y) \in \E} Q_{xy}(a) \bigg[ e^{\psi_{xy}} \times 
    \\
    &\times Z_{y}\left( t,a+\frac{g_{xy}}{t} \right) - Z_{x}(t,a)\bigg] + \frac{f_x-a}{t} \partial_a Z_{x}(t,a)
    \end{split} \, .
\end{equation}
This is the Feynman--Kac equation~\cite{kac1949on-distributions,fitzsimmons1999kacs,chetrite2015nonequilibrium} for the SIJP, to be solved with the terminal condition $Z_{x}(T,a) = 1$.

\paragraph*{\bf \em WKBJ Ansatz.} 
The self-interacting nature of the dynamics suggests that a convenient representation to solve the Feynman--Kac equation is given by the following exponential form of the MGF~\cite{assaf2017wkb-theory}
\begin{equation}
    \label{eq:WKBJ}
    Z_{x}(\zeta,\psi;t,a) = r_{x}(\zeta,\psi;t,a) e^{ \alpha(\zeta,\psi;t,a)} \, ,
\end{equation}
where we assume
\begin{equation}
    \label{eq:AsympExpWKBJ}
    \alpha(\zeta,\psi;t,a) = T u(s,a;\zeta,\psi) + o(T) \;\; \text{for} \; s = t/T \, ,
\end{equation}
with $\alpha(t,a)$ and $r(t,a)$ sufficiently smooth for the derivation below. This corresponds to a large deviation principle (LDP) with speed $T$ for the joint observable $(L(T),\Phi(T))$ where $u(s,a)$ is a time-dependent scaled cumulant generating function (SCGF). As the memory variable only enters exponentially in the action, this separates its contribution from that of the instantaneous state and assigns to it a dominant, persistent role in determining long-time fluctuations, in a manner reminiscent of an \textit{adiabatic} approximation.

By substituting the Ansatz \eqref{eq:WKBJ} into the Feynman--Kac equation \eqref{eq:FKOh}, and defining the rescaled conjugate momentum
\begin{equation}
\label{eq:ConjugateAction}
p \coloneqq t^{-1} \partial_a \alpha(\zeta,\psi;t,a),
\end{equation}
we obtain, at leading order, the eigenvalue equation
\begin{equation}
\label{eq:EigenvalueEq}
\mu(\zeta,\psi;a,p) \, r(\zeta,\psi;a,p) = \mathcal Q(\zeta,\psi;a,p) \, r(\zeta,\psi;a,p) \, ,
\end{equation}
where
\begin{equation}
\label{eq:Eigenvalue}
\mu(\zeta,\psi;a,p) \coloneqq - \partial_t \alpha(\zeta,\psi;t,a) + a p(\zeta,\psi;t,a) \, ,
\end{equation}
where, with abuse of notation, the explicit time dependence in $r$ is replaced by a dependence over $p$.
The tilted operator is defined as
\begin{equation}
\label{eq:TiltedOperator}
\mathcal Q_{xy}(\zeta,\psi;a,p) =
\begin{cases}
Q_{xy}(a)e^{\psi_{xy}+p g_{xy}},
& (x,y) \in \E \\
Q_{xx}(a)+\zeta_x+p f_x,
& x=y \\
0, &\text{otherwise}\, .
\end{cases}
\end{equation}
Here, $\mu(a,p)$ (where we omit the dependence on the counting fields) denotes the dominant eigenvalue of $\mathcal Q(a,p)$, and $r(a,p)$ its corresponding positive right eigenvector. Since the exponential factor is strictly positive in \er{eq:TiltedOperator}, the tilted generator remains irreducible on the graph $\mathcal{E}$ for every fixed value of $a$ and $p$. Moreover, if $p$ belongs to a compact set, along with $\psi$ and $\zeta$, then uniform positivity of the original rates transfers to the tilted rates and as a consequence the spectral gap of the tilted operator is uniformly bounded away from zero.

\paragraph*{\bf \em Hamilton--Jacobi framework.} In \er{eq:EigenvalueEq}, counting fields $\zeta$ and $\psi$ are supplemented by the memory $a$ and its conjugate variable $p$. The latter shifts the diagonal terms by $fp$ and exponentially reweights the off-diagonal terms by $gp$, reflecting the non-Markovian nature of the SIJP. Combining \era{eq:ConjugateAction}{eq:Eigenvalue}, the action satisfies the Hamilton--Jacobi equation
\begin{equation}
\label{eq:HamiltonJacobi}
\partial_t \alpha(t,a)
+ 
H\left(t,a,\partial_a\alpha(t,a)\right)
= 0 \, ,
\end{equation}
with time-dependent Hamiltonian
\begin{equation}
\label{eq:Hamiltonian}
H \left(t,a,P \right) \coloneqq
\mu\left(a,\frac{P}{t}\right) - a \frac{P}{t} \, ,
\end{equation}
and canonical momentum
\begin{equation}
P \coloneqq \partial_a\alpha(t,a)=tp \, .
\end{equation}
Hamilton's equations are
\begin{align}
\dot a &= \partial_P H = \frac{1}{t}
\left[ \partial_p\mu(a,p)-a \right] \\
\dot P &= -\partial_a H = -\partial_a\mu(a,p)+p \, .
\end{align}
Since $P=tp$ and $\dot P=p+t\dot p$, the characteristic equations in the original variables become
\begin{eqnarray}
\label{eq:HamiltonEqs1}
t\dot a &=& \partial_p\mu(a,p)-a \\
\label{eq:HamiltonEqs2}
t\dot p &=&  - \partial_a\mu(a,p) \, .
\end{eqnarray}

Combining the eigenvalue problem \eqref{eq:EigenvalueEq} with Eqs.\ \eqref{eq:HamiltonEqs1}--\eqref{eq:HamiltonEqs2}, subject to $(a_0,p_T)=(0,0)$, provides a practical route to the large deviations. For fixed $(\zeta,\psi)$, one solves the $d\times d$ eigenvalue problem over a grid in $(a,p)$---or, more generally, over a grid of dimension twice that of the memory---and then integrates the characteristics across this spectral landscape. The resulting characteristic trajectory $(a_t,p_t) \coloneqq (a_t,p(t,a_t))$ selects the solution manifold governing the typical temporal evolution of the memory and its conjugate variable associated with the rare fluctuation of interest, a feature absent from the long-time large-deviation theory of Markov processes~\cite{touchette2009the-large,jack2010large,chetrite2015nonequilibrium}.

Finally, by integrating the Legendre transform of the Hamiltonian \eqref{eq:Hamiltonian}, we obtain the SCGF
\begin{equation}
    \label{eq:SCGFFinal}
    u(\zeta, \psi; 0,0) = - T^{-1} \int_0^T \left[ t p_t  \dot{a}_t - \mu(a_t,p_t) + a_t p_t \right] \, dt \, .
\end{equation}
If multiple characteristic branches satisfy the boundary conditions $(a_0,p_T)=(0,0)$ the relevant one is that yielding the largest SCGF above. The notation $u(0,0)$ highlights that the expression is integrated backward from a terminal condition $\alpha(T,a) = 0$. 
Finally, via G\"{a}rtner--Ellis theorem, it is possible to extract the convex envelop $I^*$ of the true large deviation rate function $I$ as the Legendre--Fenchel transform of the SCGF \eqref{eq:SCGFFinal}, i.e.,
\begin{equation}
    \label{eq:LegendreFenchel}
    I^*(\ell,\phi) = \sup_{\zeta,\psi} 
    \left[
        \zeta \ell + \psi \phi - u(0,0;\zeta,\psi)
    \right]    
    \, ,
\end{equation}
where $I^* \leq I$ and
\begin{equation}
    \label{eq:ProbaRate}
    \mathbb{P}[L(T)=\ell,\Phi(T)=\phi] \asymp e^{- T I(\ell,\phi)} \, ,
\end{equation}
for $\sum_{x \in \S} \ell_x = 1$, $\ell_x \geq 0$, $\phi_{xy} \geq 0$, and $\sum_{y:(x,y) \in \E} \phi_{xy} = \sum_{y:(y,x) \in \E} \phi_{yx}$.

\Era{eq:EigenvalueEq}{eq:TiltedOperator}, together with \era{eq:HamiltonEqs1}{eq:HamiltonEqs2} and the SCGF \eqref{eq:SCGFFinal}, constitute the main original contribution of this paper, extending the spectral theory of large deviations for occupation measures and fluxes to non-Markovian SIJPs. 
We finally remark that although the Hamilton--Jacobi framework is more commonly associated with weak-noise large deviations~\cite{freidlin1984random,assaf2017wkb-theory,grafke2019numerical,Chabane2022}, its emergence here can be understood as a consequence of the adiabatic Ansatz \eqref{eq:WKBJ}, which effectively introduces a separation between fast state dynamics and slow evolution of the memory.

\paragraph*{\bf \em Adiabatic Doob process.} So far, we have characterised rare fluctuations of SIJPs through the SCGF and rate function. We now construct a stochastic dynamics in which these fluctuations become typical. For time-homogeneous Markov processes this is achieved through the generalised Doob transform (or driven or auxiliary process)~\cite{jack2010large,jack2015effective,chetrite2013nonequilibrium,chetrite2015nonequilibrium,garrahan2016classical}. For SIJPs, the construction is complicated by the time-dependent generator $Q(A_t)$ and by the feedback of the conditioning on the empirical memory. At leading order, however, the WKBJ Ansatz \eqref{eq:WKBJ} reduces the global conditioning to a local spectral problem parametrised by the slowly evolving pair $(a_t,p_t)$.

At bulk times $t=O(T)$, $Q(a_t)$ and the tilted operator $\mathcal Q(a_t,p_t)$ are approximately frozen over intermediate windows of length $\tau$ such that
\begin{equation}
    \Gamma(a,p)^{-1}\ll\tau\ll T,
\end{equation}
where $\Gamma(a,p)$ is the spectral gap of $\mathcal Q(a,p)$. Irreducibility and positivity of the allowed rates ensure a simple principal eigenvalue $\mu(a,p)$ with positive left and right eigenvectors $l(a,p)$ and $r(a,p)$. If the optimal characteristic remains in a compact region, cf.\ discussion below \er{eq:TiltedOperator}, $\Gamma(a,p)$ stays bounded away from zero. The configurational dynamics therefore relaxes within each window while the spectral data adiabatically follow $(a_t,p_t)$.

The corresponding Doob generator then reads
\begin{equation}
    \label{eq:DrivenOperator}
    Q^{\mathrm{dr}}_{xy}(t)
    =
    r_x(a_t,p_t)^{-1}
    \mathcal Q_{xy}(a_t,p_t)
    r_y(a_t,p_t)
    -
    \mu(a_t,p_t)\delta_{xy} \, .
\end{equation}
This is a probability conserving generator with frozen invariant density $l_x(a_t,p_t)r_x(a_t,p_t)$. The doob process is time-inhomogeneous through $(a_t,p_t)$ and provides the leading-order adiabatic generalisation of the Doob transform to SIJPs.

\paragraph*{\bf \em Self-induced East model.}
We illustrate our results with a self-interacting extension of the East model~\cite{jackle1991a-hierarchically}. We consider a one-dimensional lattice $X(t)=x(t)=(x_0(t),\ldots,x_{N-1}(t))$, with $x_i\in\{0,1\}$ and a fixed facilitating boundary $x_0(t)=1$ for $t \geq 0$ \cite{banuls2019using,Monthus2022}. Denoting by $x^{(i)}$ the configuration obtained from $x$ by flipping spin $i$, the allowed transitions have rates
\begin{equation}
    \label{eq:SelfInducedEast}
    Q_{xy}(A_t)
    =
    \sum_{i=1}^{N-1}
    2 x_{i-1}\,\omega(A_t)\,
    \mathbf{1}_{x^{(i)}}(y) .
\end{equation}
\Er{eq:SelfInducedEast} implements the East model constraint of a spin being able to flip only if its preceding neigbour is up, but where the rate $\omega$ is memory-dependent through the empirical activity per site $A_t= K_t/(Nt)$, where $K_t$ is the total number of spin flips up to time $t$. For this rate we choose the functional form 
\begin{equation}
    \label{eq:CubicRateEast}
    \omega(a)
    =
    a-
    \frac{(a-\alpha)(a-\beta)(a-\gamma)}
    {a^2+\delta^2},
\end{equation}
with $0<\alpha<\beta<\gamma$. Writing
$\sigma_1=\alpha+\beta+\gamma$,
$\sigma_2=\alpha\beta+\alpha\gamma+\beta\gamma$, and
$\sigma_3=\alpha\beta\gamma$, the condition
\begin{equation}
    |\delta^2-\sigma_2|
    <
    2\sqrt{\sigma_1\sigma_3}
\end{equation}
ensures $\omega(a)>0$. For fixed $a$, the symmetric flip rates yield an average of $N/2$ facilitated sites giving the self-consistency map $F(a)=\omega(a)$. The typical activity therefore has fixed points $a= \alpha,\beta,\gamma$, with $\alpha$ and $\gamma$ stable, and $\beta$ unstable, producing a bistable self-induced dynamics as in~\cite{coghi2026proof}.

\begin{figure}[t!]
    \centering
    \includegraphics[width=\columnwidth,keepaspectratio=true]{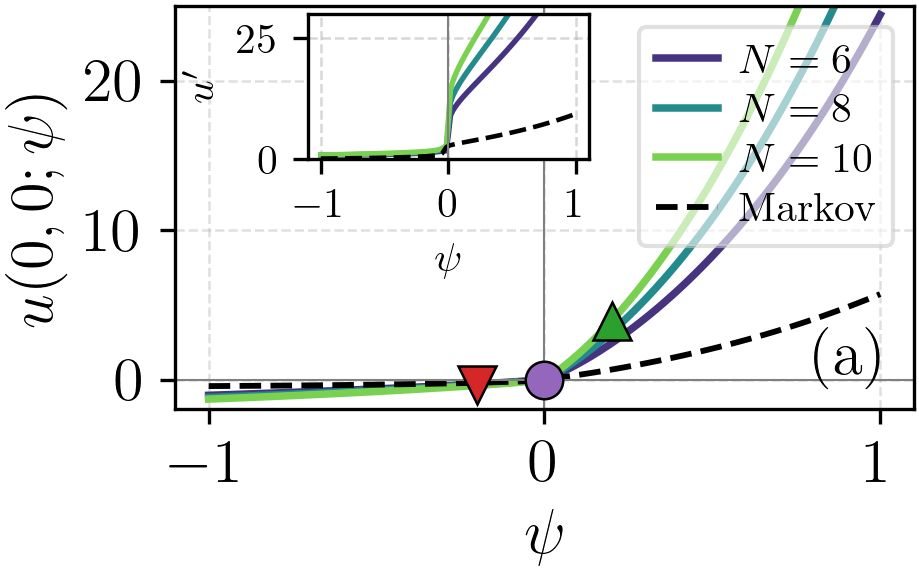} 
    \includegraphics[width=\columnwidth,keepaspectratio=true]{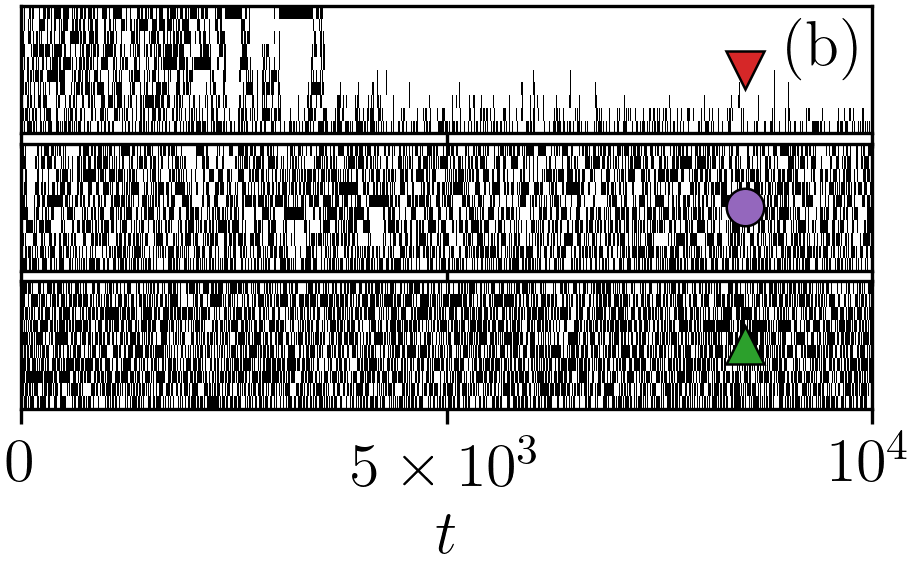}
       
    \caption{(a) SCGF of the activity for the self-induced East model at $N=6,8,10$, compared with the Markov East model at $N=10$. Inset: derivative of the SCGF. The increasingly sharp change of slope with $N$ signals an active--inactive dynamical transition, strongly enhanced by the self-induced feedback. (b) Representative trajectories of the generalised Doob process for $\psi=-0.2,0,0.2$. Negative bias suppresses activity and progressively freezes the dynamics, while positive bias enhances the number of spin flips. The untilted trajectory corresponds to typical dynamics of the self-induced East model.}
    \label{fig:SCGF}
\end{figure}

We study fluctuations of the activity by setting $f=\zeta=0$, $g=1/N$, and $\psi_{xy}\equiv\psi$ on all allowed transitions. We diagonalise the tilted operator \eqref{eq:TiltedOperator} on a grid in $(a,p)$, integrate the characteristics \eqref{eq:HamiltonEqs1}--\eqref{eq:HamiltonEqs2}, and obtain the SCGF from \er{eq:SCGFFinal}. Figure~\ref{fig:SCGF}(a) shows the resulting SCGF for $N=6,8,10$, together with the standard Markovian East model at $N=10$. The self-induced dynamics displays an increasingly sharp active--inactive crossover with increasing $N$, substantially sharper than the Markovian benchmark. This is particularly evident in the inset, where the rapid change in the slope of the SCGF provides a finite-size signature of an emerging dynamical phase transition reinforced by the memory feedback.

The corresponding Doob dynamics makes the temporal structure of the fluctuations explicit by allowing efficient simulation of the rare trajectories. Figure~\ref{fig:SCGF}(b) shows representative trajectories for $\psi=-0.2,0,0.2$. The case $\psi=0$ (central panel) shows a typical trajectory of the  self-induced East model. A negative tilt (top panel for $\psi=-0.2$) has the effect of suppressing the activity, leading to an inactive dynamical state: the suppression becomes progressively stronger as the trajectory evolves, the slow memory feeds the reduced activity back into the driven rates, making subsequent flips increasingly unlikely and eventually producing a frozen bubble of down spins. A positive tilt (bottom panel for $\psi=0.2$) in contrast enhances the overall number of spin flips and leads to an active dynamical phase. 

\paragraph*{\bf \em Conclusions.} 
Here we presented a general formalism to compute the dynamical large deviations of self-interacting jump processes in the generating function picture. To our knowledge this is the first extension of the spectral approach based on titled generators to non-Markovian systems. Our central result is the derivation of the relevant tilted operators coupled via Hamilton--Jacobi equations which encode the separation of timescales inherent to self-interaction. Furthermore, the adiabatic Doob process construction provides efficient access to the trajectories realising rare fluctuations. While the main text discussed the continuous-time case, the 
\hyperref[appendix:Notation]{End Matter}  
provides the equivalent formulation for discrete-time dynamics.

By reducing the computationally demanding part of the problem to the diagonalisation of a tilted operator, the framework opens the way to studying systems that are currently inaccessible to direct variational approaches. 
We can see direct application of our general framework to biologically inspired autochemotactic models~\cite{gelimson2016multicellular,hokmabad2022chemotactic}. Interesting follow ups to this work would be to generalise these methods to open quantum systems with memory~\cite{carollo2021large}, and to connect to learning and adaptive decision-making~\cite{sutton2018reinforcement}.

\paragraph*{\bf \em Acknowledgments.}

F.C.\ is supported by a Leverhulme Early Career Fellowship No.\ ECF-2025-482. 
J.P.G.\ acknowledges support from EPSRC grant no.\ EP/V031201/1 and The Leverhulme Trust Grant no.\ RPG-2024-112.

\paragraph*{\bf \em End Matter: Notation.} 
\label{appendix:Notation}
We use upper case for random variables (or functions) and lower case for their realisations, while reserving $\mathbb{P}$ for probability: $\mathbb{P}(X=x)$ means the probability of the random variable $X$ taking value $x$, for which we often just write $\mathbb{P}(x)$. 
We use a compact notation for multidimensional objects: given vector $\pi \coloneqq (\pi_x)_{x=1:d}$ and matrix $Q \coloneqq (Q_{xy})_{x,y=1:d}$, $\pi Q$ denotes their internal product, $(\pi Q)_y = \sum_{x=1}^d \pi_x Q_{xy}$, and use $\circ$ for Hadamard product, with $\pi \circ Q$ having elements $\pi_x Q_{xy}$. 
For scalar quantities we write the time dependence with a subscript, e.g.\ $A_t$, for compactness.
$\mathbf{1}_x(\cdot)$ is the indicator function, which gives one if the argument coincides with $x$ or zero otherwise. 
We use overbar to mark typical values, e.g., $\lim_{t \rightarrow \infty} A_t = \bar{a}$. 
The symbol $\asymp$ indicates equality up to sub-exponential factors in a large parameter, so that $\mathbb{P}(\cdot) \asymp e^{-T I(\cdot)}$ is equivalent to $I(\cdot) = \lim_{T \to \infty} T^{-1} \log \mathbb{P}(\cdot)$. 
We use the ``maths'' convention of probabilities being row vectors, stochastic generators having rows adding up to zero, and time-propagation being left to right, e.g.\ $\partial_t \mu = \mu Q$.

\paragraph*{\bf \em End-matter: Self-interacting Markov chains.}
\label{appendix:SIMCs}

We consider the SIMC  $(X_t)_{0 \leq t \leq T}$ of discrete configurations, $x \in \mathcal{S}$, where we inherit the same notation used for the state space and allowed transitions of SIJPs. At each discrete time $t \in \mathbb{N}$, the SIMC jumps from $x$ to $y$, such that $(x,y) \in \E$ according to the row-normalised transition matrix 
\begin{equation}
    \label{eq:TransMatSIMC}
    M_{xy}(a) \coloneqq \mathbb{P}[X_{t+1}=y|X_{t}=x, A_t = a] \, ,
\end{equation}
which depends on the previous outcome of the empirical observable
\begin{equation}
    \label{eq:EmpObsSIMC}
    A_t = \frac{1}{t} \sum_{t'=1}^{t} f_{X_{t'}} + \frac{1}{t} \sum_{t'=1}^{t} g_{X_{t'-1},X_{t'}} \, ,
\end{equation}
where, for simplicity, we set $f_{X_0} = 0$.

Let the pair-empirical occupation measure between $(x,y) \in \mathcal{E}$ in the time window $[t,T]$ be defined as
\begin{equation}
    \label{eq:PairEmpSIMC}
    \Phi_{xy}(t,T) = \frac{1}{T-t} \sum_{t'=1}^{T-t} \mathbf{1}_x(X_{t+t'-1}) \mathbf{1}_y(X_{t+t'}) \, ,
\end{equation}
then the MGF takes the form
\begin{equation}
    \label{eq:MomentSIMC}
    Z_{x}(t,a;\psi) \coloneqq \mathbb{E} \left[ e^{(T-t) B_{t,T}} | X_t = x, A_t = a \right] \, ,
\end{equation}
where 
\begin{equation}
    \label{eq:ObservableSIMC}
    B_{t,T} \coloneqq 
    \psi \, \Phi \, .
\end{equation}
Notice that \er{eq:PairEmpSIMC} already includes the (one-time) empirical occupation measure by marginalisation--a feature of discrete-time chains.

We follow a derivation similar to the one presented above. Initially, we write $(T-t) B_{t,T} = B_t^{t+1} + (T-t-1)B_{t+1}^T$. Then, we use conditioning and the Markov property to express $Z_{x}(t,a)$ as a function of itself at the next time step. The following resulting equation
\begin{equation}   
    \label{eq:FKSIMC}
    \begin{split}
    Z_{x}(t,a;\psi) &= \sum_{y \neq x} M_{xy}(a)  e^{\psi_{xy}} \times \\ 
    &\hspace{-0.5cm} \times Z_{y}\left( t+1,\frac{t}{t+1} a + \frac{1}{t+1} (g_{xy} + f_y);\psi \right)
    \, ,
    \end{split}
\end{equation}
is a recursive relation; it is the discrete-time analogous of the generalised Feynman--Kac equation \eqref{eq:FKOh}.

It is convenient already at this stage to assume the WKBJ Ansatz
\begin{equation}
    \label{eq:AnsatzExpSIMC}
    Z_{x}(t,a;\psi) = r_{x}(t,a;\psi) e^{\alpha(t,a;\psi)} \, ,
\end{equation}
with the asymptotic expansion in \er{eq:AsympExpWKBJ}. By further denoting $\Delta_{xy} = g_{xy} + f_y$, introducing the rescaled moment variable \eqref{eq:ConjugateAction}, and substituting \er{eq:AnsatzExpSIMC} in \eqref{eq:FKSIMC}, we obtain the eigenvalue equation
\begin{equation}
    \mu(a,p;\psi) \, r_{x}(t,a;\psi) = \mathcal{M}_{xy}(a,p;\psi) \, r_{x} (t,a;\psi) \, ,
\end{equation}
where eigenvalue and tilted operator are, respectively,
\begin{align}
    \mu(a,p) &= e^{- \partial_t \alpha(t,a) + p \, a} \\
    \mathcal{M}(a,p) &= M(a) e^{\psi + p \Delta} \, .
\end{align}

The Hamilton--Jacobi structure that follows from the WKBJ Ansatz yields the following characteristic equations
\begin{align}
    \label{eq:CharacteristicsSIMC}
    t \dot{a} &= \partial_p \log \mu(a,p) - a \\
    t \dot{p} &= - \partial_a \log \mu(a,p) \, ,
\end{align}
whose solution $(a_t,p_t)$ allows us to derive the SCGF
\begin{equation}
    \label{eq:FinalSCGFSIMC}
    u(\psi; 0,0) = - T^{-1}\int_0^T( t p_t \dot{a}_t - \log \mu(a_t,p_t) + a_t p_t ) \, dt \, .
\end{equation}
Consequently, the convex envelope of the true rate function can be extracted via G\"{a}rtner--Ellis theorem as
\begin{equation}
    \label{eq:FinalRateSIMC}
    I^*(\phi) = \sup_{\psi} \left[ \psi \phi - u(0,0;\psi) \right] \, .
\end{equation}

Finally, the leading-order adiabatic generalised Doob process is obtained by tracing the characteristics $(a_t,p_t)$, which is itself solution of the system \eqref{eq:CharacteristicsSIMC}. Therefore, the Doob process transition matrix is
\begin{equation}
    \label{eq:DoobSIMC}
    \mathcal{M}^{\mathrm{dr}}_{xy}(t)
    =
    (\mu(a_t,p_t)r_x(a_t,p_t))^{-1}
    M_{xy}(a_t,p_t)
    r_y(a_t,p_t) \, .
\end{equation}

\bibliography{bibliography-06092026}

\bibliographystyle{apsrev4-2}

\end{document}